\documentclass[iop]{emulateapj}
\usepackage{amsmath}
\usepackage{times}
\usepackage{graphics, subfigure}
\usepackage{epsfig}
\usepackage{multirow}
\usepackage{ulem}
\usepackage{pdfpages}

\def\simlt{\mathrel{\hbox{\rlap{\hbox{\lower4pt\hbox{$\sim$}}}\hbox{$<$}}}}
\def\simgt{\mathrel{\hbox{\rlap{\hbox{\lower4pt\hbox{$\sim$}}}\hbox{$>$}}}}
\newcommand{\Mshell}{M_{\textrm{shell}}}

\begin{document}

\title{The progenitors of core-collapse supernovae suggest thermonuclear origin for the explosions}

\author{Doron Kushnir\altaffilmark{1}} \altaffiltext{1}{Institute for Advanced Study, Einstein Drive, Princeton, NJ, 08540, USA, kushnir@ias.edu}

\begin{abstract}
Core-collapse supernovae (CCSNe) are the explosions of massive stars following the collapse of the stars' iron cores. Poznanski (2013) has recently suggested an observational correlation between the ejecta velocities and the inferred masses of the red supergiant progenitors of type II-P explosions, which implies that the kinetic energy of the ejecta ($E_{\textrm{kin}}$) increases with the mass of the progenitor. I point out that the same conclusion can be reached from the model-free observed correlation between the ejected $^{56}$Ni masses ($M_{\textrm{Ni}}$) and the luminosities of the progenitors for type II supernovae, which was reported by Fraser et al. (2011). This correlation is in an agreement with the predictions of the collapse-induced thermonuclear explosions (CITE) for CCSNe and in a possible contradiction with the predictions of the neutrino mechanism. I show that a correlation between $M_{\textrm{Ni}}$ and $E_{\textrm{kin}}$ holds for all types of CCSNe (including type Ibc). This correlation suggests a common mechanism for all CCSNe, which is predicted for CITE, but is not produced by current simulations of the neutrino mechanism. Furthermore, the typical values of $E_{\textrm{kin}}$ and $M_{\textrm{Ni}}$ for type Ibc explosions are larger by an order of a magnitude than the typical values for II-P explosions, a fact which disfavors progenitors with the same initial mass range for these explosions. Instead, the progenitors of type Ibc explosions could be massive Wolf-Rayet stars, which are predicted to yield strong explosions with low ejecta masses (as observed) according to CITE. In this case, there is no deficit of high mass progenitors for CCSNe, which was suggested under the assumption of a similar mass range for the progenitors of types II-P and Ibc supernovae.     
\end{abstract}

% -------------------------- End of abstract -----------------------


\section{Introduction}
\label{sec:Introduction}

There is strong evidence that supernovae of types II and Ibc are explosions of massive stars \citep[e.g.][]{87a,Arnett1989,vanDyk1992,smartt2009}, involving the collapse of the stars' iron cores and ejection of the outer layers. It is widely thought that the observed $\sim10^{51}\,\textrm{erg}$ kinetic energy of the ejecta ($E_{\textrm{kin}}$) is due to the deposition of a small fraction ($\sim1\%$) of the gravitational energy ($\sim10^{53}\,\textrm{erg}$) released in neutrinos \citep[see][for reviews]{Bethe90,Janka2012}. So far, this scenario has not been demonstrated from first principles. In fact, one-dimensional simulations indicate that the neutrinos do not deposit sufficient energy. While some explosions were obtained in multi-dimensional simulations with simplified neutrino transport, the fundamental mechanism would only be satisfactorily demonstrated once accurate three-dimensional simulations, with all relevant physical process taken into account, become available. \citet[][]{BBFH} suggested a different mechanism for the explosion during core-collapse that does not involve the emitted neutrinos. In this proposed scenario, increased burning rates due to adiabatic heating of the outer shells as they collapse lead to a thermonuclear explosion \citep[see also][]{Hoyle60,Fowler64}. This collapse-induced thermonuclear explosion (CITE) naturally produces $\sim10^{51}\,\textrm{erg}$ from thermonuclear burning of $\sim 1\,M_{\odot}$ (gain of $\sim \textrm{MeV}/m_{p}$).

\citet[][]{Kushnir14} have shown that CITE is possible in some (tuned) one-dimensional initial profiles, which include shells of mixed helium and oxygen, but resulting in weak explosions, $\lesssim10^{50}\,\textrm{erg}$, and negligible amounts of $^{56}$Ni are ejected. In \citet{Kushnir2015} I have recently used two-dimensional simulations of rotating massive stars to explore the conditions required for CITE to operate successfully. I found out that for stellar cores that include slowly (a few percent of breakup) rotating $\sim0.1-10\,M_{\odot}$ explosive shells of He-O with densities of $\textrm{few}\times10^{3}\,\textrm{g}\,\textrm{cm}^{-3}$, an ignition of a thermonuclear detonation that unbinds the stars' outer layers is obtained. With a series of simulations that cover a wide range of the progenitor masses and profiles, I showed that CITE is insensitive to the assumed profiles and thus a robust process that leads to supernova explosions for rotating massive stars. The resulting explosions have $E_{\textrm{kin}}$ in the range of $10^{49}-10^{52}\,\textrm{erg}$, and ejected $^{56}$Ni masses ($M_{\textrm{Ni}}$) of up to $\sim1\,M_{\odot}$, both of which cover the observed ranges of core-collapse supernovae (CCSNe, including types II and Ibc). CITE predicts that stronger explosions (i.e., larger $E_{\textrm{kin}}$ and higher $M_{\textrm{Ni}}$) are from progenitors with higher masses. Testing if the required initial conditions for CITE to operate exist in nature is difficult observationally, but here I show observational evidence from CCSNe that are in agreement with the prediction that stronger explosions are from progenitors with higher masses, which implies that CITE may be the dominant mechanism for CCSNe explosions.

In recent years, direct identifications of the progenitors have been made for CCSNe in pre-explosion images, and they provide powerful tests for  CCSNe theories \citep[e.g.][]{smartt2009,Leonard2011,Smartt2015}. Several observed correlations between the properties of the progenitors and the supernovae suggest that more massive progenitors lead to stronger explosions. \citet{Poznanski2013} has recently suggested an observational correlation between the ejecta velocities and the inferred masses of the red supergiant progenitors of type II-P explosions. The correlation implies that $E_{\textrm{kin}}$ is approximately proportional to the mass of the progenitor cubed. \citet{Poznanski2013} suggested that the same correlation can be also deduced for type II-P supernovae from the observed uniformity of the light-curve plateau duration \citep[][]{Poznanski2009,Arcavi2012} and the correlation between the light-curve luminosities and ejecta velocities \citep[][]{Hamuy2002,Nugent2006}. In Section~\ref{sec:Ni-L}, I point out that more massive progenitors leading to stronger explosions can be deduced in a model-independent way from the observed correlation between $M_{\textrm{Ni}}$ and the progenitor luminosities. This observed correlation was first reported by \citet{Fraser2011} \citep[see also a closely related correlation between $M_{\textrm{Ni}}$ and the masses of the progenitors, suggested by][]{Smartt2009b}. Unlike progenitor masses or $E_{\textrm{kin}}$, whose inferences rely upon models (massive star evolution models or complicated light-curve models, respectively) and thus are subjective to large systematics uncertainties due to model assumptions, both $M_{\textrm{Ni}}$ and progenitor luminosities are model-free and can be directly derived from observations. Furthermore, these two quantities can be deduced for all type II explosions and are not restricted to type II-P supernovae. I reproduce the correlation between $M_{\textrm{Ni}}$ and the progenitor luminosities with an updated data (Section~\ref{sec:Ni-L observations}) and show that it is in an agreement with the predictions of CITE (Section~\ref{sec:Ni-L CITE}) and in a possible contradiction with the predictions of the neutrino mechanism (Section~\ref{sec:Ni-L nu}).

The use of $M_{\textrm{Ni}}$ as an indicator for $E_{\textrm{kin}}$ is based on an observed correlation between $M_{\textrm{Ni}}$ and $E_{\textrm{kin}}$ shown in Section~\ref{sec:Ni-Ekin}. I demonstrate that this correlation holds for all types of CCSNe, including both types II and Ibc (Section~\ref{sec:Ni-Ekin observations}). This universal correlation suggests a common explosion mechanism for all CCSNe, which is predicted for CITE, but it is not produced by current simulations of the neutrino mechanism (Section~\ref{sec:Ni-Ekin common}). Furthermore, the typical values of $E_{\textrm{kin}}$ and $M_{\textrm{Ni}}$ for type Ibc explosions are larger by an order of a magnitude from the typical values for type II-P explosions. This fact disfavors a similar mass range for the progenitors of these events, and suggests that the progenitors of type Ibc explosions are massive Wolf-Rayet (WR) stars (Section~\ref{sec:Ni-L Ibc}). Since WR stars have more massive cores and stripped envelopes, CITE predicts that they lead to stronger explosions and relatively low ejecta masses, both of which are consistent with observations. Progenitor studies that assume a similar mass range for the progenitors of type II-P and type Ibc supernovae suggest a deficit of high mass progenitors $(\simgt20\,M_{\odot}$) for CCSNe, and if true, it would imply that higher mass stars produce ``failed supernovae'' -- weak explosions that are very faint \citep[e.g., see][]{smartt2009}. However, If massive WR stars are the progenitors of type Ibc supernovae, there is no deficit of high mass progenitors for CCSNe \citep[][]{Smartt2015}.
 
\section{Ejected $^{56}$Ni masses versus the luminosities of the progenitor}
\label{sec:Ni-L}

\subsection{Observations}
\label{sec:Ni-L observations}

\begin{figure}
        \subfigure[]{
             \includegraphics[width=0.9\textwidth]{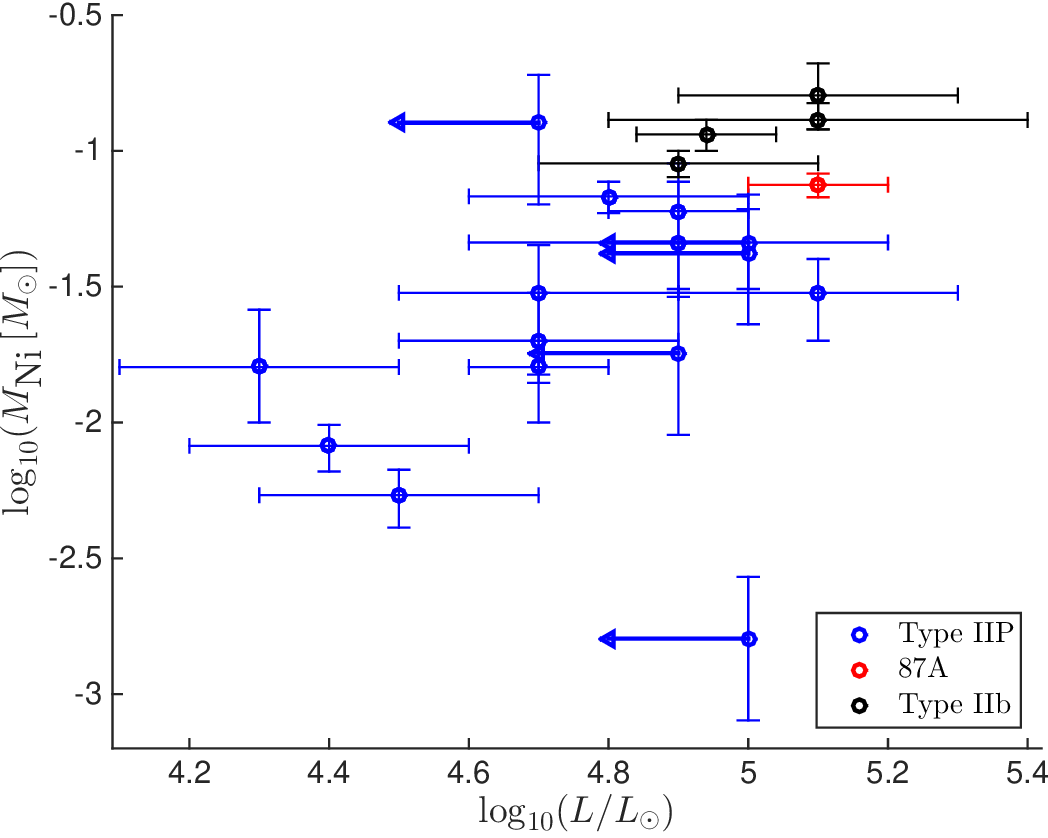}
        }
        \subfigure[]{
             \includegraphics[width=0.45\textwidth]{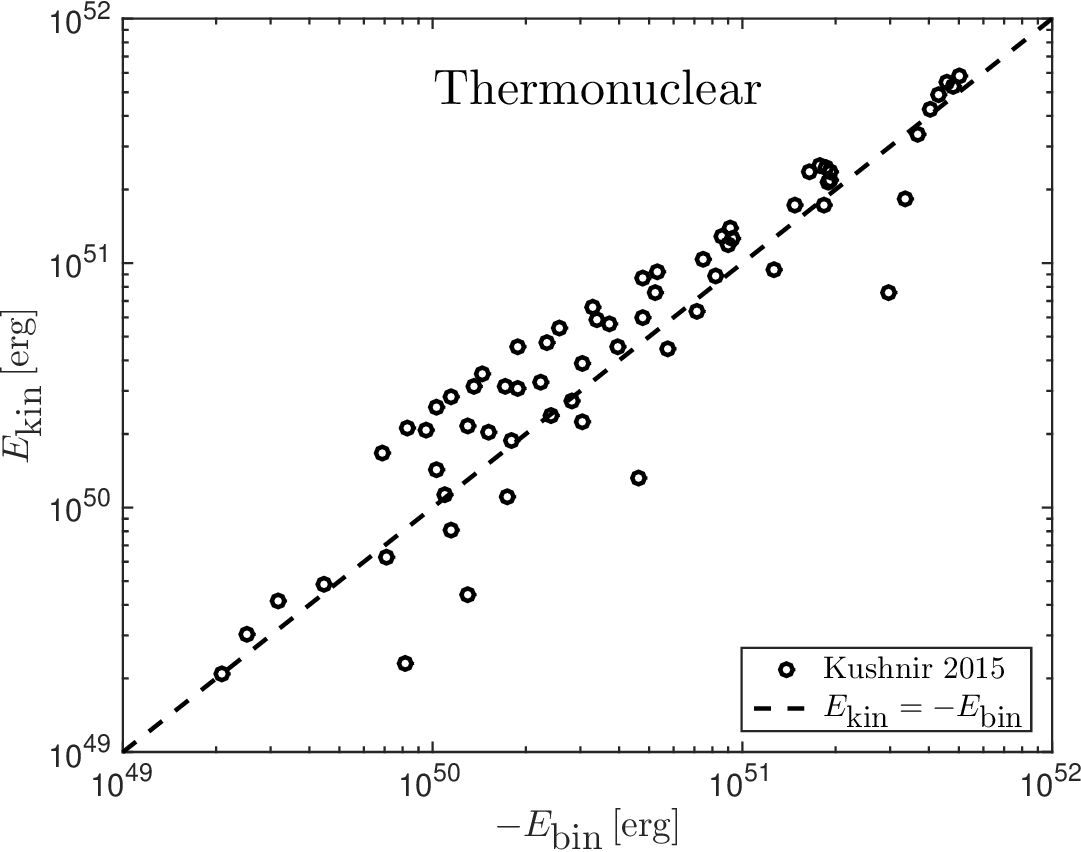}
        }
        \subfigure[]{
             \includegraphics[width=0.45\textwidth]{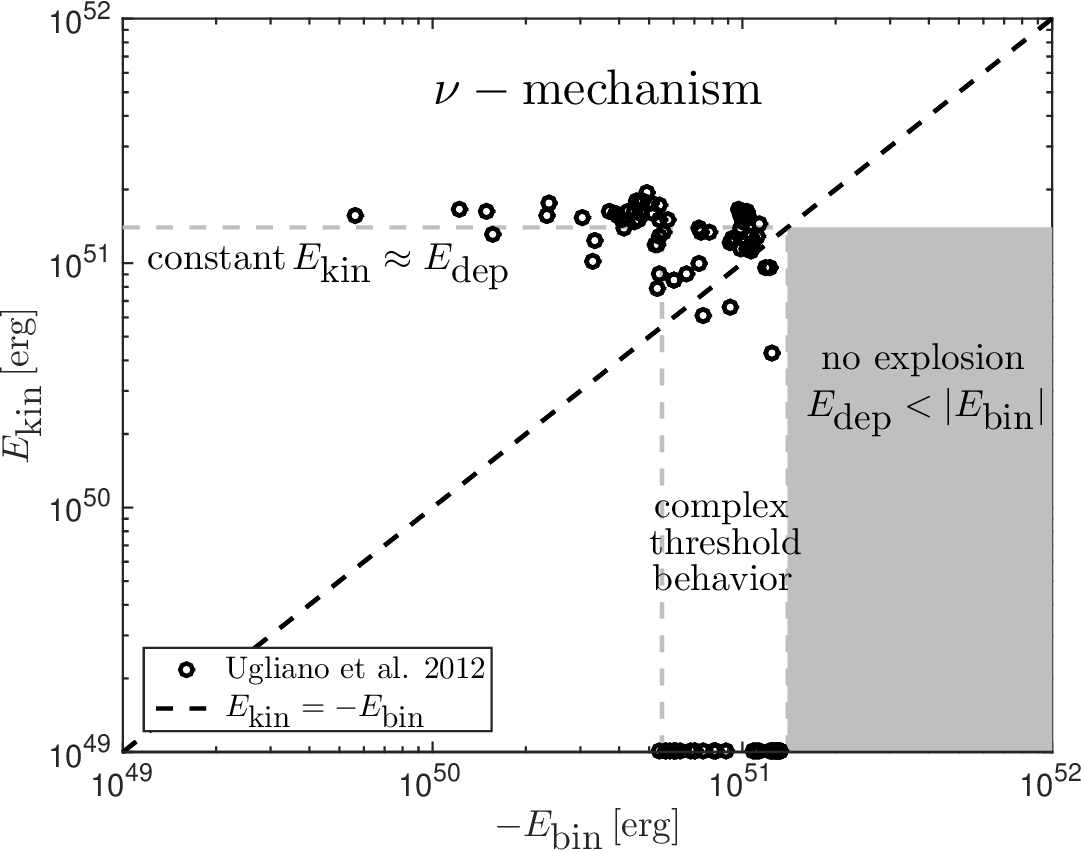}
        }
      \caption{Panel (a): The observed correlation between $M_{\textrm{Ni}}$ and the luminosities of the progenitors for type II supernovae, which was first reported by \citet{Fraser2011}, is reproduced here with updated data. The sample includes all supernovae from \citet{Smartt2015}, for which an estimate of $M_{\textrm{Ni}}$ is available in the literature, supplemented with SN 1987A and SN 1993J (see Table~\ref{tbl:Ni-L}). In the cases that $M_{\textrm{Ni}}$ lacks an error estimate, an error of $50\%$ was assumed ($10\%$ for SN 1987A). More luminous progenitors eject larger masses of $^{56}$Ni. Since more luminous progenitors are more massive (with more negative binding energy, $E_{\textrm{bin}}$) and since larger values of $M_{\textrm{Ni}}$ imply larger $E_{\textrm{kin}}$ (see Section~\ref{sec:Ni-Ekin} and Figure~\ref{fig:Ni-Ekin}), the correlation implies that more massive progenitors lead to stronger explosions. The range of $M_{\textrm{Ni}}$ roughly corresponds to $E_{\textrm{kin}}\sim\textrm{few}\times10^{50}-\textrm{few}\times10^{51}\,\textrm{erg}$. Panel (b): The kinetic energy of the ejecta as function of $E_{\textrm{bin}}$ at the base of the explosive shell for the CITE simulations that exploded successfully from \citet{Kushnir2015}. Panel (c):  The kinetic energy of the ejecta as function of $E_{\textrm{bin}}$ exterior to the iron core for the neutrino mechanism simulations of \citet[][]{Ugliano2012}. The points at $10^{49}\,\textrm{erg}$ represent failed explosions. 
   \label{fig:Ni-L}}
\end{figure}

The observed correlation between $M_{\textrm{Ni}}$ and the luminosities of the progenitors, which was reported by \citet{Fraser2011}, is reproduced with updated data in panel (a) of Figure~\ref{fig:Ni-L}. The sample includes all supernovae from \citet{Smartt2015}, for which estimates of $M_{\textrm{Ni}}$ are available in the literature, supplemented with SN 1987A and SN 1993J (see Table~\ref{tbl:Ni-L}). A clear correlation over one order of magnitude for both $M_{\textrm{Ni}}$ and for the luminosity of the progenitor is apparent, where the range of $M_{\textrm{Ni}}$ roughly corresponds to $E_{\textrm{kin}}\sim\textrm{few}\times10^{50}-\textrm{few}\times10^{51}\,\textrm{erg}$ (see Figure~\ref{fig:Ni-Ekin}). More luminous progenitors eject larger masses of $^{56}$Ni. Note that SN 1987A and type IIb supernovae have the largest progenitors luminosities and the largest $M_{\textrm{Ni}}$ values, a property that will be discussed in Section~\ref{sec:Ni-Ekin}. 

\begin{figure}
\includegraphics[width=0.9\textwidth]{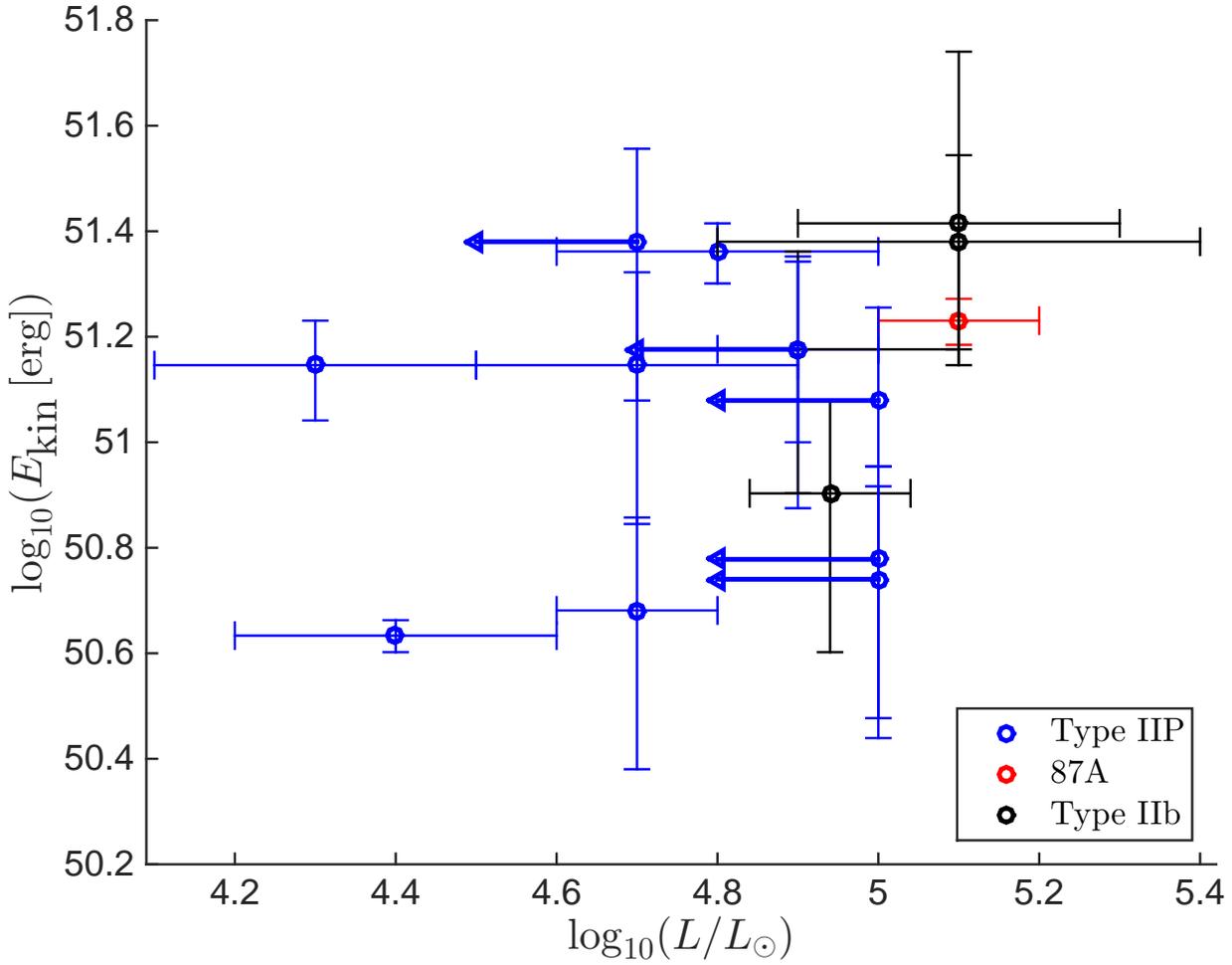}
\caption{The observed correlation between the estimated $E_{\textrm{kin}}$ and the luminosities of the progenitors. The sample includes all supernovae from \citet{Smartt2015}, for which an estimate of $E_{\textrm{kin}}$ is available in the literature, supplemented with SN 1987A and SN 1993J (see Table~\ref{tbl:Ni-L}). In the cases that $E_{\textrm{kin}}$ lacks an error estimate, an error of $50\%$ was assumed ($10\%$ for SN 1987A). The estimates of $E_{\textrm{kin}}$ from observations involve complicated light-curve modeling (which can include large systematic uncertainties). This is probably the reason for the weak observed correlation that is obtained when using $E_{\textrm{kin}}$ compared to the strong observed correlation that is obtained when using $M_{\textrm{Ni}}$ (panel (a) of Figure~\ref{fig:Ni-L}), which is model-free and can be directly derived from observations.
\label{fig:Ekin-L}}
\end{figure}

\begin{figure}
\includegraphics[width=0.9\textwidth]{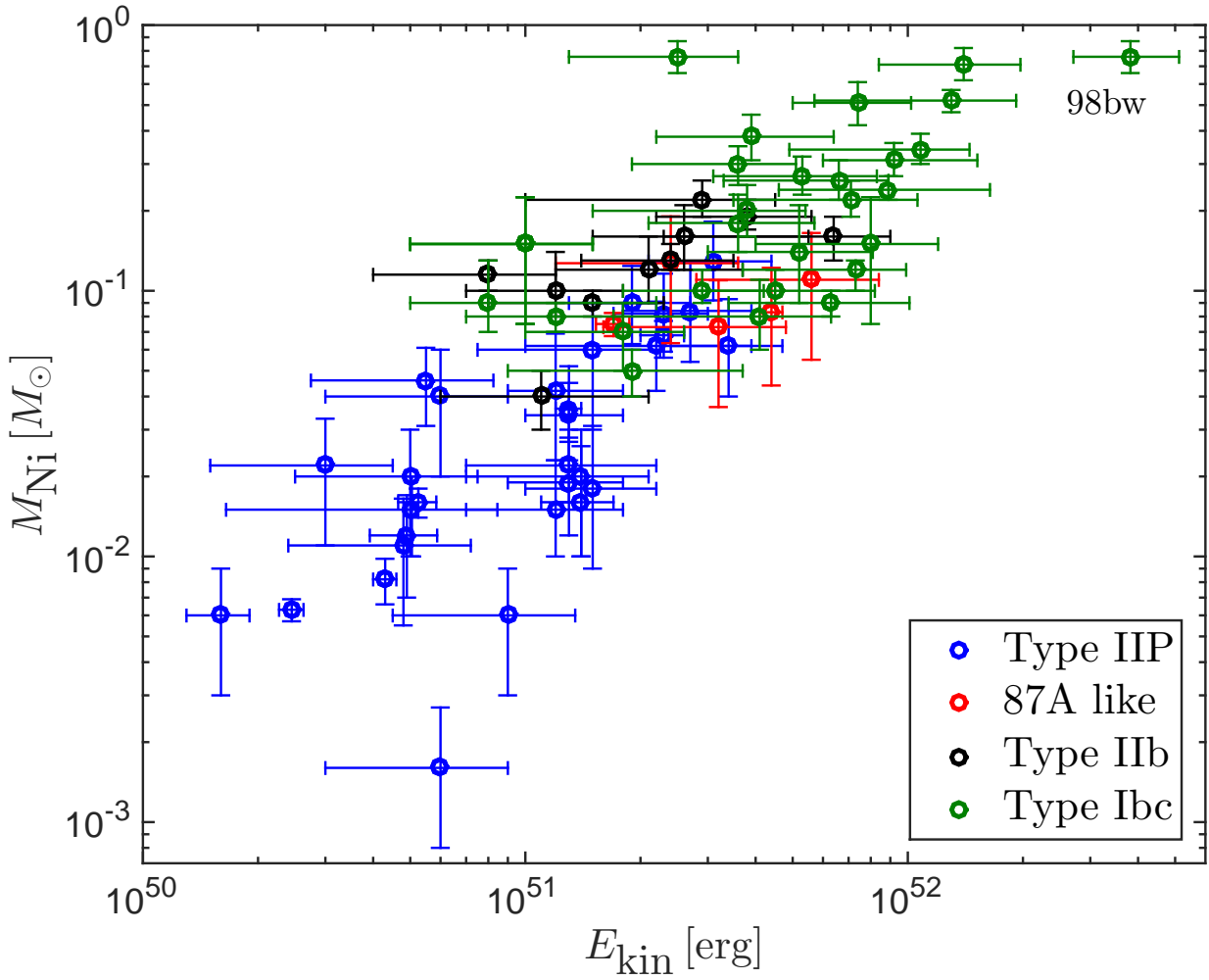}
\caption{Estimates of $E_{\textrm{kin}}$ and $M_{\textrm{Ni}}$ from the literature for $70$ observed supernovae (see Table~\ref{tbl:observations}). This is the same compilation of \citet{Kushnir2015} with a few more events. In the case that $E_{\textrm{kin}}$ or $M_{\textrm{Ni}}$ lack an error estimate, an error of $50\%$ was assumed ($10\%$ for SN 1987A). The estimates of $E_{\textrm{kin}}$ from observations involve complicated light-curve modeling (which can include large systematic uncertainties). However, unlike the situation in Figure~\ref{fig:Ekin-L}, in this case the sample is large and it spans more than two orders of magnitude in $M_{\textrm{Ni}}$ and $E_{\textrm{kin}}$, such that the large systematic uncertainties are less important.
\label{fig:Ni-Ekin}}
\end{figure}

It would seem natural to inspect the correlation between $E_{\textrm{kin}}$ and the luminosity of the progenitors, rather than using $M_{\textrm{Ni}}$ as an indicator for $E_{\textrm{kin}}$. The main motivation against using the inferred $E_{\textrm{kin}}$ from observations is the complicated light-curve modeling that is involved for its estimation (which can include large systematic uncertainties), compared with the model-free determination of $M_{\textrm{Ni}}$. Indeed, only a weak correlation is obtained between the estimated $E_{\textrm{kin}}$ reported in the literature (see Table~\ref{tbl:Ni-L}) and the luminosity of the progenitors, as shown in Figure~\ref{fig:Ekin-L}. The advantage of using $M_{\textrm{Ni}}$ over $E_{\textrm{kin}}$ is evident by comparing panel (a) of Figure~\ref{fig:Ni-L} to Figure~\ref{fig:Ekin-L}. The correlation between $M_{\textrm{Ni}}$ and $E_{\textrm{kin}}$ (Section~\ref{sec:Ni-Ekin}) that is the justification for using $M_{\textrm{Ni}}$ as an indicator for $E_{\textrm{kin}}$ suffers as well from the large systematic uncertainties in the estimation of $E_{\textrm{kin}}$. However, in this case the sample is large and it spans more than two orders of magnitude in $M_{\textrm{Ni}}$ and $E_{\textrm{kin}}$, such that the large systematic uncertainties are less important.

Since more luminous progenitors are more massive and since larger values of $M_{\textrm{Ni}}$ imply larger $E_{\textrm{kin}}$, the correlation between $M_{\textrm{Ni}}$ and the luminosities of the progenitors implies that more massive progenitors lead to stronger explosions, the same qualitative result found by \citet{Poznanski2013}. The model-free measurements of $M_{\textrm{Ni}}$ and of the luminosities of the progenitors are more robust than the estimates of the masses of the progenitors (which depend on stellar evolution models) and of $E_{\textrm{kin}}$ (which depend on complicated light-curve modeling). Furthermore, \citet{Poznanski2013} used the Fe II $\lambda 5169$ absorption feature to estimate the velocity of the ejecta, which limits the analysis for events other than type II-P. 

\subsection{The prediction of CITE agrees with observations}
\label{sec:Ni-L CITE}

A primary prediction of CITE is that $E_{\textrm{kin}}$ increases with the mass of the progenitor \citep{Kushnir2015}. This is more apparent by considering the binding energy of the shells to be ejected, $E_{\textrm{bin}}$ (corrected for thermal energy), which is more negative for more massive progenitors. We can write quite generally that $E_{\textrm{kin}}$ is given by
\begin{equation}\label{eq:KE}
E_{\textrm{kin}}\approx E_{\textrm{dep}}+E_{\textrm{bin}},
\end{equation}
where $E_{\textrm{dep}}$ is the energy deposited in the ejecta. For CITE, the deposited energy is thermonuclear,  $E_{\textrm{dep}}\sim\Mshell\times\textrm{MeV}/m_{p}$, where $\Mshell$ is the mass of shell of the thermonuclear fuel (the explosive shell). The relevant binding energy in this case is the one exterior to the base of the explosive shell, $E_{\textrm{bin}}\sim -G M_{\textrm{base}}\Mshell/r_{\textrm{base}}$, where $M_{\textrm{base}}$ and $r_{\textrm{base}}$ are the enclosed mass and the radius at the base of the explosive shell, respectively. $E_{\textrm{dep}}$ and $|E_{\textrm{bin}}|$ are comparable, since $\textrm{few}\times G M_{\textrm{base}}/r_{\textrm{base}}\approx\textrm{MeV}/m_{p}$ \citep[][]{Kushnir14}. Therefore, $E_{\textrm{kin}}$ can never exceeds significantly $|E_{\textrm{bin}}|$, and in the absence of a tuning between $E_{\textrm{dep}}$ and $E_{\textrm{bin}}$, $E_{\textrm{kin}}$ cannot be much smaller than $|E_{\textrm{bin}}|$. Therefore, $E_{\textrm{kin}}\sim|E_{\textrm{bin}}|$ for CITE. This order of magnitude estimate is validated in panel (b) of Figure~\ref{fig:Ni-L}, which shows the results of the CITE simulations that exploded successfully from \citet{Kushnir2015}. The conclusion is that the prediction of CITE agrees with the observation that more massive progenitors lead to stronger explosions.

\subsection{The prediction of the neutrino mechanism possibly contradicts observations}
\label{sec:Ni-L nu}

For the neutrino mechanism, $E_{\textrm{dep}}$ is the energy deposited by neutrinos. Since from basic considerations the iron core is similar over a wide range of progenitor masses (the iron core is approximately a Chandrasekhar-mass white dwarf), $E_{\textrm{dep}}$ is roughly constant over a wide progenitor mass range. However, the relevant binding energy in this case, the one exterior to the iron core, changes significantly between different progenitor masses. Therefore, as long as $E_{\textrm{dep}}\gg|E_{\textrm{bin}}|$, Equation~\eqref{eq:KE} predicts that $E_{\textrm{kin}}\approx E_{\textrm{dep}}\approx\textrm{constant}$. At some progenitor mass $E_{\textrm{dep}}$ is comparable to $|E_{\textrm{bin}}|$, such that for higher progenitor masses the explosion fails, since the deposited energy by neutrinos is smaller than the (absolute) binding energy. This behavior should be general for the neutrino mechanism, and probably does not depend on the specific scenario in which the star explodes. In fact, this behavior should hold for every scenario in which the deposited energy is dominated by the stellar core and is not sensitive to the binding energy of the shells to be ejected. So we expect $E_{\textrm{kin}}$ to be constant up to some value of $|E_{\textrm{bin}}|$ (threshold progenitor mass) and then to rapidly fall to zero (failed explosions). 

The results of \citet[][]{Ugliano2012} for the neutrino mechanism are shown in Panel (c) of Figure~\ref{fig:Ni-L}. I use the values of $E_{\textrm{bin}}$, as reported by \citet[][]{Ugliano2012}, which are defined exterior to the iron core (at a mass coordinate of $\approx1.5\,M_{\odot}$), and are approximately the binding energies of the shells that are to be ejected. At low progenitor masses (low $|E_{\textrm{bin}}|$) the value of $E_{\textrm{kin}}$ is indeed constant. However, instead of a sharp drop for $E_{\textrm{kin}}$ at some value of $|E_{\textrm{bin}}|$, there is a complicated behavior near $|E_{\textrm{bin}}|\approx10^{51}\,\textrm{erg}$, which received much attention recently \citep[][]{O'Connor2011,Ugliano2012,Pejcha2015,Ertl2015}. The range of binding energies over which this complicated behavior is obtained is only a factor of $\approx2$ and is of no importance for the current discussion. Another complication in the behavior for the neutrino mechanism is the predicted weak explosions ($\approx10^{50}\,\textrm{erg}$) for the lowest mass progenitors \cite[electron-capture supernova (ECSN);][]{Nomoto1984,Nomoto1987,Kitaura2006,Janka2008,Wanajo2011}. However, the combination of two different mechanisms (iron core-collapse at high progenitor masses and electron-capture at low progenitor masses) is not supported by the uniformity of the observed correlations for the entire progenitor mass range (see the discussion at the end of Section~\ref{sec:Ni-Ekin common}). In summary, the prediction of the neutrino mechanism is a roughly constant $E_{\textrm{kin}}$ for a wide range of progenitor masses and a sharp drop (maybe with a complicated behavior over a small range of progenitor masses) at some progenitor mass. This is in a possible contradiction with the observation that more massive progenitors lead to stronger explosions. It is yet to be seen whether accurate three-dimensional simulations of the neutrino mechanism, with all relevant physical process taken into account, would reproduce this observation.

\section{Ejected $^{56}$Ni masses versus the kinetic energies of the ejecta}
\label{sec:Ni-Ekin}

\subsection{Observations}
\label{sec:Ni-Ekin observations}

Estimates of $E_{\textrm{kin}}$ and $M_{\textrm{Ni}}$ for $70$ observed supernovae within comoving radial distance of $<100\,\textrm{Mpc}$ (to exclude rare events) are listed in Table~\ref{tbl:observations} and are shown in Figure~\ref{fig:Ni-Ekin}. This is the same compilation of \citet{Kushnir2015} with a few more events. Note that the distribution of the sample in the $E_{\textrm{kin}}$--$M_{\textrm{Ni}}$ plane does not represent the relative rates of the events. A clear correlation over two orders of magnitude for both $E_{\textrm{kin}}$ and $M_{\textrm{Ni}}$ is apparent. Stronger explosions eject larger masses of $^{56}$Ni. This correlation allowed the use of  $M_{\textrm{Ni}}$ as an indicator for $E_{\textrm{kin}}$ in Section~\ref{sec:Ni-L}. The estimates of $E_{\textrm{kin}}$ from observations involve complicated light-curve modeling (which can include large systematic uncertainties). However, unlike the situation in Figure~\ref{fig:Ekin-L}, in this case the sample is large and it spans more than two orders of magnitude in $M_{\textrm{Ni}}$ and $E_{\textrm{kin}}$, such that the large systematic uncertainties are less important.

\subsection{A common mechanism for all CCSNe}
\label{sec:Ni-Ekin common}

The correlation between $E_{\textrm{kin}}$ and $M_{\textrm{Ni}}$ holds for all types of CCSNe (types II and Ibc), and spans the entire observed ranges of $E_{\textrm{kin}}$ ($\sim10^{50}-10^{52}\,\textrm{erg}$) and $M_{\textrm{Ni}}$ ($\sim10^{-3}-1\,M_{\odot}$). This correlation suggests a common mechanism for all CCSNe, from the weakest observed explosions to the strongest ones. Such a common mechanism is predicted for CITE \citep{Kushnir2015}, but seems unlikely for the neutrino mechanism, for two reasons. The first reason is that current simulations of the neutrino mechanism do not produce strong ($\sim10^{52}\,\textrm{erg}$) explosions \citep[see the discussion in][]{Janka2012}. The second reason is that weak ($\sim10^{50}\,\textrm{erg}$) explosions would require an extreme tuning for the neutrino mechanism. In the case that $|E_{\textrm{bin}}|\sim10^{51}\,\textrm{erg}$, the fraction of the gravitational energy ($\sim10^{53}\,\textrm{erg}$) released in neutrinos that is deposited should be $\sim2\%$ for moderate ($\sim10^{51}\,\textrm{erg}$) explosions, and should be $\sim1.1\%$ for weak explosions (a tuning of $\sim10^{-3}$). In the case that $|E_{\textrm{bin}}|\sim10^{50}\,\textrm{erg}$, the fraction of the gravitational energy released in neutrinos that is deposited should be $\sim0.2\%$ for weak explosions (again, a tuning of $\sim10^{-3}$). The possibility that a different mechanism (ECSN) is operating for the lowest mass progenitors is not supported by the smooth observed correlations, which suggest a common mechanism for all CCSNe. This is demonstrated more robustly by the correlation between $M_{\textrm{Ni}}$ and the V-band plateau luminosities, which suggests a common mechanism for weak and moderate events \citep[][Figure 16 there]{Spiro2014}.

\subsection{The progenitors of type Ibc explosions are massive Wolf-Rayet (WR) stars -- no deficit of high mass progenitors for CCSNe}
\label{sec:Ni-L Ibc}

The distribution of the different types of events in the $E_{\textrm{kin}}$--$M_{\textrm{Ni}}$ plane indicates that the sequence II-P, 87A like, IIb, Ibc is a sequence of $E_{\textrm{kin}}$ and of $M_{\textrm{Ni}}$ (this sequence is evident even when considering only $M_{\textrm{Ni}}$, which is more robustly observed). 

$E_{\textrm{kin}}$ and $M_{\textrm{Ni}}$ for type Ibc explosions are larger by an order of a magnitude than $E_{\textrm{kin}}$ and $M_{\textrm{Ni}}$ for type II-P explosions, respectively. Let us consider the possibility that the progenitors of types Ibc and type II-P supernovae have a similar mass range, and that the different display of the supernova is solely because of the stripping of the hydrogen envelope for the type Ibc case. One expects that in this case $E_{\textrm{kin}}$ and $M_{\textrm{Ni}}$ would be similar for types II-P and Ibc, since these parameters are determined by the explosion mechanism, which takes place at the interior of the star, and is independent of the hydrogen envelope properties (and whether it exists or not). However, as pointed out above, $E_{\textrm{kin}}$ and $M_{\textrm{Ni}}$ for type Ibc explosions are larger by an order of a magnitude than the typical values for type II-P explosions. Therefore, the possibility that the progenitors of types Ibc and type II-P supernovae have a similar mass range is disfavored by observations. 

One caveat is that $^{56}$Ni-powered events like type Ibc are hard to find when $M_{\textrm{Ni}}$ is small, while type II events that initially powered by shock cooling can be observed even if they produce no $^{56}$Ni at all. Therefore, the lack of type Ibc events with small values of $M_{\textrm{Ni}}$ may be because of a selection bias. One possible way to check for such a bias is to calculate the (Pearson) partial correlation between $\log_{10}(E_{\textrm{kin}})$ and $\log_{10}(M_{\textrm{Ni}})$ given the distances to the events, which is  $\rho\simeq0.73$ with a $p$-value of $\simeq1.4\cdot10^{-13}$, suggesting that such a bias is unlikely.

The second discussed possibility for the progenitors of type Ibc are Wolf-Rayet (WR) stars \citep[see also the suggestion that the observed progenitor of the type Ib SN PTF13bvn is a WR star;][]{Cao2013,Groh2013}. Since these stars have more massive cores, CITE predicts that they lead to stronger explosions and larger amounts of $^{56}$Ni are ejected. This continues the trend that was established in Section~\ref{sec:Ni-L} for type II explosions, that more massive progenitors yield stronger explosions. One argument given by \citet{Bersten2014} and by \citet{Smartt2015} against WR stars being the progenitors of type Ibc is the low estimated mass of the ejecta (typicaly $1-4\,M_{\odot}$) compared to the mass of WR stars (typicaly $8-20\,M_{\odot}$). However, this is a problem only if one assumes that most of the mass of the progenitor is ejected, as predicted by the neutrino mechanism. For CITE, only the mass exterior to the base of the explosive shell is ejected, and in the case that there is no hydrogen envelope, this mass agrees with the estimated ejected mass from observations \citep{Kushnir2015}. It is further predicted by CITE for WR progenitors that the interior mass to the base of the explosive shell collapses and forms a massive black hole. So, assuming CITE explosions, strong type Ibc explosions with low ejecta masses are consistent with massive WR progenitors. Progenitor studies that assume a similar mass range for the progenitors of types II-P and Ibc supernovae suggest a deficit of high mass progenitors ($\simgt20\,M_{\odot}$) for CCSNe \citep[e.g., see][]{smartt2009}. However, If massive WR stars are the progenitors of type Ibc supernovae, there is no deficit of high mass progenitors for CCSNe \citep[][]{Smartt2015}.

In summary, the observational evidence suggests that the sequence II-P, 87A like, IIb, Ibc is a progenitor mass sequence, where more massive progenitors lead to stronger explosions.

\acknowledgments I thank Subo Dong, Avishay Gal-Yam, Boaz Katz and Eran Ofek for useful discussions and for a thorough reading of the manuscript. D.~K. gratefully acknowledges support from the Friends of the Institute for Advanced Study. 

%-----------------------------------------------------------------------------
% --------------------------      BIBLIOGRAPGHY ---------------------------
%-----------------------------------------------------------------------------

\bibliographystyle{apj}

\begin{thebibliography}{99}
\bibitem[Arcavi et al.(2012)]{Arcavi2012} Arcavi, I., Gal-Yam, A., Cenko, S.~B., et al.\ 2012, \apjl, 756, L30
\bibitem[Arnett et al.(1989)]{Arnett1989} Arnett, W.~D., Bahcall, J.~N., Kirshner, R.~P., \& Woosley, S.~E.\ 1989, \araa, 27, 629
\bibitem[Bersten et al.(2014)]{Bersten2014} Bersten, M.~C., Benvenuto, O.~G., Folatelli, G., et al.\ 2014, \aj, 148, 68
\bibitem[Bethe(1990)]{Bethe90} Bethe, H.~A.\ 1990, Reviews of Modern Physics, 62, 801
%\bibitem[Bruenn et al.(2014)]{Bruenn2014} Bruenn, S.~W., Lentz, E.~J., Hix, W.~R., et al.\ 2014, arXiv:1409.5779
\bibitem[Burbidge et al.(1957)]{BBFH} Burbidge, E.~M.,
Burbidge, G.~R., Fowler, W.~A., \& Hoyle, F.\ 1957, Reviews of Modern Physics, 29, 547
\bibitem[Cao et al.(2013)]{Cao2013} Cao, Y., Kasliwal, M.~M., Arcavi, I., et al.\ 2013, \apjl, 775, L7
\bibitem[Dall'Ora et al.(2014)]{Dall'Ora2014} Dall'Ora, M., Botticella, M.~T., Pumo, M.~L., et al.\ 2014, \apj, 787, 139
\bibitem[Ertl et al.(2015)]{Ertl2015} Ertl, T., Janka, H.-T., Woosley, S.~E., Sukhbold, T., \& Ugliano, M.\ 2015, arXiv:1503.07522 
\bibitem[Fowler \& Hoyle(1964)]{Fowler64} Fowler, W.~A., \& Hoyle, F.\ 1964, \apjs, 9, 201
\bibitem[Fraser et al.(2011)]{Fraser2011} Fraser, M., Ergon, M., Eldridge, J.~J., et al.\ 2011, \mnras, 417, 1417
\bibitem[Groh et al.(2013)]{Groh2013} Groh, J.~H., Georgy, C., \& Ekstr{\"o}m, S.\ 2013, \aap, 558, L1
\bibitem[Hamuy \& Pinto(2002)]{Hamuy2002} Hamuy, M., \& Pinto, P.~A.\ 2002, \apjl, 566, L63
\bibitem[Hamuy(2003)]{Hamuy2003} Hamuy, M.\ 2003, \apj, 582, 905
\bibitem[Hendry et al.(2005)]{Hendry2005} Hendry, M.~A., Smartt, S.~J., Maund, J.~R., et al.\ 2005, \mnras, 359, 906
\bibitem[Hendry et al.(2006)]{Hendry2006} Hendry, M.~A., Smartt, S.~J., Crockett, R.~M., et al.\ 2006, \mnras, 369, 1303
\bibitem[Hirata et al.(1987)]{87a} Hirata, K., Kajita, T., Koshiba, M., Nakahata, M.,\& Oyama, Y.\ 1987, Physical Review Letters, 58, 1490
\bibitem[Hoyle \& Fowler(1960)]{Hoyle60} Hoyle, F., \& Fowler, W.~A.\ 1960, \apj, 132, 565
\bibitem[Huang et al.(2015)]{Huang2015} Huang, F., Wang, X., Zhang, J., et al.\ 2015, arXiv:1504.00446
\bibitem[Inserra et al.(2011)]{Inserra2011} Inserra, C., Turatto, M., Pastorello, A., et al.\ 2011, \mnras, 417, 261
\bibitem[Janka et al.(2008)]{Janka2008} Janka, H.-T., M{\"u}ller, B., Kitaura, F.~S., \& Buras, R.\ 2008, \aap, 485, 199
\bibitem[Janka(2012)]{Janka2012} Janka, H.-T.\ 2012, Annual Review of Nuclear and Particle Science, 62, 407
\bibitem[Jerkstrand et al.(2015)]{Jerkstrand2015} Jerkstrand, A., Smartt, S.~J., Sollerman, J., et al.\ 2015, \mnras, 448, 2482
\bibitem[Kitaura et al.(2006)]{Kitaura2006} Kitaura, F.~S., Janka, H.-T., \& Hillebrandt, W.\ 2006, \aap, 450, 345
\bibitem[Kushnir \& Katz(2014)]{Kushnir14} Kushnir, D., \& Katz, B.\ 2014, arXiv:1412.1096
\bibitem[Kushnir(2015)]{Kushnir2015} Kushnir, D.\ 2015, arXiv:1502.03111 
\bibitem[Leonard(2011)]{Leonard2011} Leonard, D.~C.\ 2011, \apss, 336, 117 
\bibitem[Lyman et al.(2014)]{Lyman2014} Lyman, J., Bersier, D., James, P., et al.\ 2014, arXiv:1406.3667
\bibitem[Maund et al.(2004)]{Maund2004} Maund, J.~R., Smartt, S.~J., Kudritzki, R.~P., Podsiadlowski, P., \& Gilmore, G.~F.\ 2004, \nat, 427, 129 
\bibitem[Morales-Garoffolo et al.(2014)]{MG2014} Morales-Garoffolo, A., Elias-Rosa, N., Benetti, S., et al.\ 2014, \mnras, 445, 1647
\bibitem[Nomoto(1984)]{Nomoto1984} Nomoto, K.\ 1984, \apj, 277, 791
\bibitem[Nomoto(1987)]{Nomoto1987} Nomoto, K.\ 1987, \apj, 322, 206 
\bibitem[Nugent et al.(2006)]{Nugent2006} Nugent, P., Sullivan, M., Ellis, R., et al.\ 2006, \apj, 645, 841
\bibitem[O'Connor \& Ott(2011)]{O'Connor2011} O'Connor, E., \& Ott, C.~D.\ 2011, \apj, 730, 70
\bibitem[Pastorello et al.(2005)]{Pastorello2005} Pastorello, A., Baron, E., Branch, D., et al.\ 2005, \mnras, 360, 950
\bibitem[Pastorello et al.(2012)]{Pastorello2012} Pastorello, A., Pumo, M.~L., Navasardyan, H., et al.\ 2012, \aap, 537, AA141
\bibitem[Pejcha \& Thompson(2015)]{Pejcha2015} Pejcha, O., \& Thompson, T.~A.\ 2015, \apj, 801, 90
\bibitem[Poznanski et al.(2009)]{Poznanski2009} Poznanski, D., Butler, N., Filippenko, A.~V., et al.\ 2009, \apj, 694, 1067
\bibitem[Poznanski(2013)]{Poznanski2013} Poznanski, D.\ 2013, \mnras, 436, 3224
\bibitem[Smartt(2009)]{smartt2009} Smartt, S.~J.\ 2009, \araa, 47, 63
\bibitem[Smartt et al.(2009)]{Smartt2009b} Smartt, S.~J., Eldridge, J.~J., Crockett, R.~M., \& Maund, J.~R.\ 2009, \mnras, 395, 1409
\bibitem[Smartt(2015)]{Smartt2015} Smartt, S.~J.\ 2015, PASA, 32, e016
\bibitem[Spiro et al.(2014)]{Spiro2014} Spiro, S., Pastorello, A., Pumo, M.~L., et al.\ 2014, \mnras, 439, 2873
\bibitem[Taddia et al.(2012)]{Taddia2012} Taddia, F., Stritzinger, M.~D., Sollerman, J., et al.\ 2012, \aap, 537, AA140
\bibitem[Tak{\'a}ts et al.(2015)]{Takatas2015} Tak{\'a}ts, K., Pignata, G., Pumo, M.~L., et al.\ 2015, \mnras, 450, 3137
\bibitem[Tomasella et al.(2013)]{Tomasella2013} Tomasella, L., Cappellaro, E., Fraser, M., et al.\ 2013, \mnras, 434, 1636
\bibitem[Ugliano et al.(2012)]{Ugliano2012} Ugliano, M., Janka, H.-T., Marek, A., \& Arcones, A.\ 2012, \apj, 757, 69
\bibitem[Utrobin \& Chugai(2014)]{UC2014} Utrobin, V.~P., \& Chugai, N.~N.\ 2014, arXiv:1411.6480
\bibitem[van Dyk(1992)]{vanDyk1992} van Dyk, S.~D.\ 1992, \aj, 103, 1788
\bibitem[Wanajo et al.(2011)]{Wanajo2011} Wanajo, S., Janka, H.-T., M{\"u}ller, B.\ 2011, \apjl, 726, L15

\end{thebibliography}

\newpage

\newpage

\begin{deluxetable}{ccccc}
\tablecaption{The progenitors from the sample of \citet{Smartt2015}, for which estimates of $M_{\textrm{Ni}}$ are available in the literature, supplemented with SN 1987A and SN 1993J. \label{tbl:Ni-L}}
\tablewidth{0pt}
\tablehead{ \colhead{Name} & \colhead{$\log_{10}(L/L_{\odot})$} & \colhead{$E_{\textrm{kin}}\,[10^{51}\,\textrm{erg}]$} & \colhead{$^{56}\textrm{Ni mass}\,[M_{\odot}]$} & \colhead{Type}}
\startdata

03gd	&	$4.3^{+0.2}_{-0.2}$	&	$1.4^{+0.3}_{-0.3}$	&	$0.016^{+0.01}_{-0.006}$	&	IIP	\\
05cs	&	$4.4^{+0.2}_{-0.2}$	&	$0.43^{+0.03}_{-0.03}$	&	$0.0082^{+0.0016}_{-0.0016}$	&	IIP	\\
09md	&	$4.5^{+0.2}_{-0.2}$	&	--	&	$0.0054^{+0.0013}_{-0.0013}$	&	IIP	\\
06my	&	$4.7^{+0.2}_{-0.2}$	&	--	&	$0.03^{+0.015}_{-0.015}$	&	IIP	\\
12A	&	$4.7^{+0.1}_{-0.1}$	&	$0.48$	&	$0.016^{+0.002}_{-0.002}$	&	IIP	\\
13ej	&	$4.7^{+0.2}_{-0.2}$	&	$1.4^{+0.7}_{-0.7}$	&	$0.02^{+0.01}_{-0.01}$	&	IIP	\\
04et	&	$4.8^{+0.2}_{-0.2}$	&	$2.3^{+0.3}_{-0.3}$	&	$0.068^{+0.009}_{-0.009}$	&	IIP	\\
04A	&	$4.9^{+0.3}_{-0.3}$	&	--	&	$0.046^{+0.031}_{-0.017}$	&	IIP	\\
12aw	&	$4.9^{+0.1}_{-0.1}$	&	$1.5$	&	$0.06$	&	IIP	\\
12ec	&	$5.1^{+0.2}_{-0.2}$	&	--	&	$0.03^{+0.01}_{-0.01}$	&	IIP	\\
06ov	&	$<4.7$	&	$2.4$	&	$0.127$	&	IIP	\\
99gi	&	$<4.9$	&	$1.5^{+0.7}_{-0.5}$	&	$0.018^{+0.013}_{-0.009}$	&	IIP	\\
99br	&	$<5$	&	$0.6$	&	$0.0016^{+0.0011}_{-0.0008}$	&	IIP	\\
99em	&	$<5$	&	$1.2^{+0.6}_{-0.3}$	&	$0.042^{+0.027}_{-0.019}$	&	IIP	\\
09ib	&	$<5$	&	$0.55$	&	$0.046^{+0.015}_{-0.015}$	&	IIP	\\
08ax	&	$5.1^{+0.2}_{-0.2}$	&	$2.6^{+2.9}_{-1.1}$	&	$0.16^{+0.05}_{-0.04}$	&	IIb	\\
11dh	&	$4.9^{+0.2}_{-0.2}$	&	$1.5^{+0.8}_{-0.7}$	&	$0.09^{+0.01}_{-0.01}$	&	IIb	\\
13df	&	$4.94^{+0.1}_{-0.1}$	&	$0.8^{+0.4}_{-0.4}$	&	$0.115^{+0.015}_{-0.015}$	&	IIb	\\
87A	&	$5.1^{+0.1}_{-0.1}$	&	$1.7$	&	$0.075$	&	87A	\\
93J	&	$5.1^{+0.3}_{-0.3}$	&	$2.4^{+1.1}_{-1}$	&	$0.13^{+0.02}_{-0.01}$	&	IIb	\\
\enddata
\tablecomments{The luminosities of the progenitors are from \citet{Smartt2015}, except SN 1987A \citep[][]{Smartt2009b} and SN 1993J \citep[][]{Maund2004}. The estimates of $E_{\textrm{kin}}$ and of $M_{\textrm{Ni}}$ are from Table~\ref{tbl:observations}, except SN 2009md \citep[][]{Fraser2011}, SN 2006my \citep[][]{Smartt2009b}, SN 2004A \citep[][]{Hendry2006} and SN 2012ec \citep[][]{Jerkstrand2015}.}
\end{deluxetable}

\newpage

\begin{deluxetable}{cccccccccc}
\tablecaption{A compilation from the literature of estimated $E_{\textrm{kin}}$ and $M_{\textrm{Ni}}$ from the light-curves. \label{tbl:observations}}
\tablewidth{0pt}
\tablehead{ \colhead{Name} &  \colhead{$\textrm{Kinetic energy}\,[10^{51}\,\textrm{erg}]$} & \colhead{$^{56}\textrm{Ni mass}\,[M_{\odot}]$} & \colhead{Type} & \colhead{Reference} & \colhead{Name} &  \colhead{$\textrm{Kinetic energy}\,[10^{51}\,\textrm{erg}]$} & \colhead{$^{56}\textrm{Ni mass}\,[M_{\odot}]$} & \colhead{Type} & \colhead{Reference}}
\startdata

69L	&	$2.3^{+0.7}_{-0.6}$	&	$0.082^{+0.034}_{-0.026}$	&	IIP	&	3	&	73R	&	$2.7^{+1.2}_{-0.9}$	&	$0.084^{+0.044}_{-0.03}$	&	IIP	&	3	\\
83I	&	$1$	&	$0.15$	&	Ibc	&	3	&	83N	&	$1$	&	$0.15$	&	Ibc	&	3	\\
84L	&	$1$	&	$0.15$	&	Ibc	&	3	&	86L	&	$1.3^{+0.5}_{-0.3}$	&	$0.034^{+0.018}_{-0.011}$	&	IIP	&	3	\\
87A	&	$1.7$	&	$0.075$	&	87A	&	3	&	88A	&	$2.2^{+1.7}_{-1.2}$	&	$0.062^{+0.029}_{-0.02}$	&	IIP	&	3	\\
89L	&	$1.2^{+0.6}_{-0.5}$	&	$0.015^{+0.008}_{-0.005}$	&	IIP	&	3	&	90E	&	$3.4^{+1.3}_{-1}$	&	$0.062^{+0.031}_{-0.022}$	&	IIP	&	3	\\
91G	&	$1.3^{+0.9}_{-0.6}$	&	$0.022^{+0.008}_{-0.006}$	&	IIP	&	3	&	92H	&	$3.1^{+1.3}_{-1}$	&	$0.129^{+0.053}_{-0.037}$	&	IIP	&	3	\\
92ba	&	$1.3^{+0.5}_{-0.4}$	&	$0.019^{+0.009}_{-0.007}$	&	IIP	&	3	&	93J	&	$2.4^{+1.1}_{-1}$	&	$0.13^{+0.02}_{-0.01}$	&	IIb	&	11	\\
94I	&	$1.2^{+0.6}_{-0.5}$	&	$0.08^{+0.01}_{-0.01}$	&	Ibc	&	11	&	96cb	&	$2.1^{+1.6}_{-0.9}$	&	$0.12^{+0.04}_{-0.03}$	&	IIb	&	11	\\
97D	&	$0.9$	&	$0.006$	&	IIP	&	3	&	97ef	&	$8$	&	$0.15$	&	Ibc	&	3	\\
98A	&	$5.6$	&	$0.11$	&	87A	&	7	&	98bw	&	$38.2^{+13}_{-11.1}$	&	$0.76^{+0.11}_{-0.1}$	&	Ibc	&	11	\\
99br	&	$0.6$	&	$0.0016^{+0.0011}_{-0.0008}$	&	IIP	&	3	&	99cr	&	$1.9^{+0.8}_{-0.6}$	&	$0.09^{+0.034}_{-0.027}$	&	IIP	&	3	\\
99dn	&	$7.3^{+2.6}_{-3.6}$	&	$0.12^{+0.01}_{-0.02}$	&	Ibc	&	11	&	99em	&	$1.3^{+0.1}_{-0.1}$	&	$0.036^{+0.009}_{-0.009}$	&	IIP	&	1	\\
99em	&	$1.2^{+0.6}_{-0.3}$	&	$0.042^{+0.027}_{-0.019}$	&	IIP	&	3	&	99ex	&	$3.6^{+2.1}_{-1.5}$	&	$0.18^{+0.05}_{-0.04}$	&	Ibc	&	11	\\
99gi	&	$1.5^{+0.7}_{-0.5}$	&	$0.018^{+0.013}_{-0.009}$	&	IIP	&	3	&	00cb	&	$4.4^{+0.3}_{-0.3}$	&	$0.083^{+0.039}_{-0.039}$	&	87A	&	1	\\
02ap	&	$6.3^{+3.8}_{-2.9}$	&	$0.09^{+0.01}_{-0.01}$	&	Ibc	&	11	&	03Z	&	$0.245^{+0.018}_{-0.018}$	&	$0.0063^{+0.0006}_{-0.0006}$	&	IIP	&	4	\\
03bg	&	$3.8^{+1.8}_{-1.6}$	&	$0.19^{+0.03}_{-0.02}$	&	IIb	&	11	&	03gd	&	$1.4^{+0.3}_{-0.3}$	&	$0.016^{+0.01}_{-0.006}$	&	IIP	&	5	\\
03jd	&	$7.4^{+2.8}_{-2.4}$	&	$0.51^{+0.1}_{-0.09}$	&	Ibc	&	11	&	04aw	&	$6.6^{+2.3}_{-3.3}$	&	$0.26^{+0.05}_{-0.04}$	&	Ibc	&	11	\\
04dk	&	$5.3^{+3}_{-2.2}$	&	$0.27^{+0.05}_{-0.04}$	&	Ibc	&	11	&	04dn	&	$7.1^{+3.5}_{-3.6}$	&	$0.22^{+0.04}_{-0.03}$	&	Ibc	&	11	\\
04et	&	$2.3^{+0.3}_{-0.3}$	&	$0.068^{+0.009}_{-0.009}$	&	IIP	&	1	&	04fe	&	$3.6^{+1.5}_{-1.7}$	&	$0.3^{+0.05}_{-0.05}$	&	Ibc	&	11	\\
04ff	&	$2.9^{+1.6}_{-1.9}$	&	$0.22^{+0.04}_{-0.03}$	&	IIb	&	11	&	04gq	&	$5.2^{+2.9}_{-2.2}$	&	$0.14^{+0.07}_{-0.05}$	&	Ibc	&	11	\\
05az	&	$3.9^{+2.5}_{-1.7}$	&	$0.38^{+0.08}_{-0.07}$	&	Ibc	&	11	&	05bf	&	$0.8^{+1.4}_{-0.3}$	&	$0.09^{+0.04}_{-0.02}$	&	Ibc	&	11	\\
05cs	&	$0.43^{+0.03}_{-0.03}$	&	$0.0082^{+0.0016}_{-0.0016}$	&	IIP	&	1	&	05cs	&	$0.16^{+0.03}_{-0.03}$	&	$0.006^{+0.003}_{-0.003}$	&	IIP	&	2	\\
05hg	&	$2.5^{+1.1}_{-1.2}$	&	$0.76^{+0.11}_{-0.1}$	&	Ibc	&	11	&	06T	&	$1.2^{+0.6}_{-0.5}$	&	$0.1^{+0.04}_{-0.02}$	&	IIb	&	11	\\
06au	&	$3.2$	&	$0.073$	&	87A	&	8	&	06el	&	$6.4^{+2.6}_{-4.1}$	&	$0.16^{+0.03}_{-0.03}$	&	IIb	&	11	\\
06ep	&	$4.1^{+2.2}_{-2.4}$	&	$0.08^{+0.03}_{-0.02}$	&	Ibc	&	11	&	06ov	&	$2.4$	&	$0.127$	&	87A	&	8	\\
07C	&	$3.8^{+1.6}_{-2.3}$	&	$0.2^{+0.05}_{-0.04}$	&	Ibc	&	11	&	07Y	&	$1.9^{+1.8}_{-1}$	&	$0.05^{+0.01}_{-0.01}$	&	Ibc	&	11	\\
07gr	&	$2.9^{+1.3}_{-1.1}$	&	$0.1^{+0.02}_{-0.01}$	&	Ibc	&	11	&	07od	&	$0.5$	&	$0.02$	&	IIP	&	6	\\
07ru	&	$13^{+6.2}_{-7.3}$	&	$0.52^{+0.05}_{-0.05}$	&	Ibc	&	11	&	07uy	&	$10.8^{+3.7}_{-5.9}$	&	$0.34^{+0.05}_{-0.04}$	&	Ibc	&	11	\\
08D	&	$4.5^{+3.7}_{-1.7}$	&	$0.1^{+0.02}_{-0.01}$	&	Ibc	&	11	&	08ax	&	$2.6^{+2.9}_{-1.1}$	&	$0.16^{+0.05}_{-0.04}$	&	IIb	&	11	\\
08in	&	$0.505^{+0.34}_{-0.34}$	&	$0.015^{+0.005}_{-0.005}$	&	IIP	&	1	&	08in	&	$0.49^{+0.098}_{-0.098}$	&	$0.012^{+0.005}_{-0.005}$	&	IIP	&	2	\\
09E	&	$0.6$	&	$0.04$	&	IIP	&	7	&	09bb	&	$9.2^{+6}_{-3.2}$	&	$0.31^{+0.05}_{-0.04}$	&	Ibc	&	11	\\
09bw	&	$0.3$	&	$0.022$	&	IIP	&	10	&	09ib	&	$0.55$	&	$0.046^{+0.015}_{-0.015}$	&	IIP	&	12	\\
09jf	&	$8.9^{+7.5}_{-4.3}$	&	$0.24^{+0.03}_{-0.02}$	&	Ibc	&	11	&	11bm	&	$14^{+5.7}_{-5.6}$	&	$0.71^{+0.11}_{-0.09}$	&	Ibc	&	11	\\
11dh	&	$1.5^{+0.8}_{-0.7}$	&	$0.09^{+0.01}_{-0.01}$	&	IIb	&	11	&	11hs	&	$1.1^{+1}_{-0.5}$	&	$0.04^{+0.01}_{-0.01}$	&	IIb	&	11	\\
12A	&	$0.48$	&	$0.011$	&	IIP	&	9	&	12A	&	$0.525^{+0.06}_{-0.06}$	&	$0.016^{+0.002}_{-0.002}$	&	IIP	&	1	\\
12aw	&	$1.5$	&	$0.06$	&	IIP	&	10	&	13df	&	$0.8^{+0.4}_{-0.4}$	&	$0.115^{+0.015}_{-0.015}$	&	IIb	&	13	\\
13ej	&	$1.4^{+0.7}_{-0.7}$	&	$0.02^{+0.01}_{-0.01}$	&	IIP	&	14	&	iPTF13bvn	&	$1.8^{+0.8}_{-0.8}$	&	$0.07^{+0.02}_{-0.02}$	&	Ibc	&	11	\\			
\enddata
\tablecomments{REFERENCES.--(1) \citet[][]{UC2014};(2) \citet[][]{Spiro2014};(3) \citet[][]{Hamuy2003};(4) \citet[][]{Hendry2005};(5) \citet[][]{Inserra2011};(6) \citet[][]{Pastorello2012};(7) \citet[][]{Pastorello2005};(8) \citet[][]{Taddia2012};(9) \citet[][]{Tomasella2013};(10) \citet[][]{Dall'Ora2014};(11) \citet[][]{Lyman2014}; (12) \citet[][]{Takatas2015}; (13) \citet[][]{MG2014}; (14) \citet[][]{Huang2015}}
\end{deluxetable}

\end{document}